\documentclass[aps,prl,reprint]{revtex4-1}

\usepackage{amsmath}
\usepackage{amssymb}
\usepackage{graphicx}

\begin{document}

\title{Optical control of the refractive index of a single atom}

    \author{Tobias Kampschulte} \email{kampschulte@iap.uni-bonn.de}
    \author{Wolfgang Alt}
    \author{Stefan Brakhane}
    \author{Martin Eckstein}
    \author{Ren\'e Reimann}
    \author{Artur Widera}
    \author{Dieter Meschede}
    \affiliation{Institut f{\"u}r Angewandte Physik, Universit{\"a}t Bonn,
    Wegelerstra{\ss}e~8, 53115~Bonn, Germany}


\begin{abstract}
We experimentally demonstrate the elementary case of electromagnetically
induced transparency (EIT) with a single atom inside an optical cavity probed
by a weak field. We observe the modification of the dispersive and absorptive
properties of the atom by changing the frequency of a control light field.
Moreover, a strong cooling effect has been observed at two-photon resonance,
increasing the storage time of our atoms twenty-fold to about 16 seconds. Our
result points towards all-optical switching with single photons.
\end{abstract}

\maketitle

The properties of an optically probed atomic medium can be changed
dramatically by the coherent interaction with a near-resonant
control light field. The simultaneous interaction with two light
fields gives rise to intriguing phenomena, such as EIT
\cite{Har90,Fle05}, leading for example to the propagation of slow
light \cite{Har92,Xia95,Sch96,Hau99}. The mixed states of light
and matter formed in this case can be interpreted as polaritons
\cite{Fle00}, originally introduced in the many-particle limit of
solid state physics \cite{Hop58}. This concept is illustrated by
the storage of light in and retrieval from an atomic ensemble
\cite{Fle00,Phi01,Liu01,Sch09}. At the single-particle level, the
control of optical properties can be utilized for atom-light
quantum interfaces \cite{Boo07} or quantum gates \cite{Tur95}.
Coherent population trapping or EIT in systems like single ions in
free space \cite{Slo10} or superconducting artificial atoms
\cite{Kel10,Abd10}, have recently been observed. With atoms, one
usually requires large ensembles that are optically thick for the
probe field in order to obtain strong effects. The coupling to a
high-finesse cavity can be utilized to observe strong EIT signals
also with single particles. Whereas this has been achieved in the
resonant regime \cite{Muc10}, we study the off-resonant case,
where we are sensitive to both absorptive and dispersive effects
and observe significant cooling.

\begin{figure}[b]
    \includegraphics[width=0.3\textwidth]{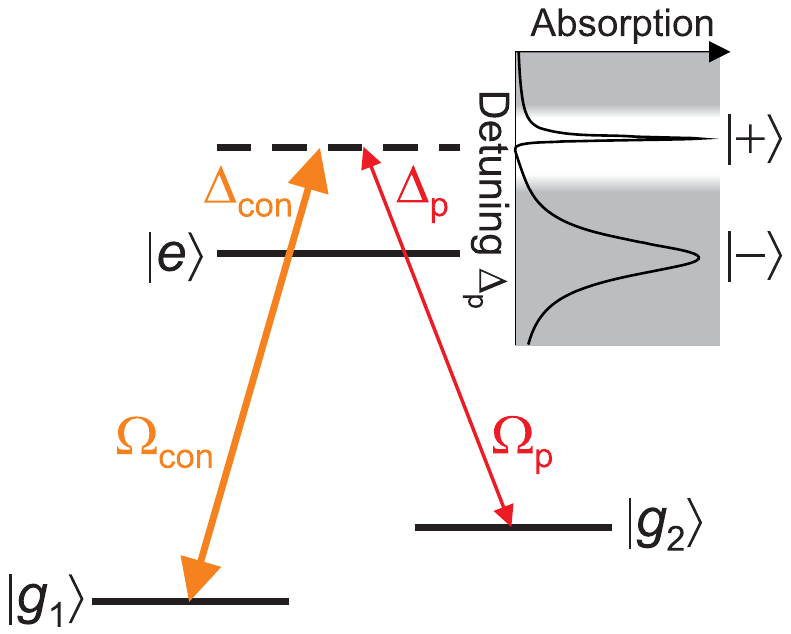}
    \caption{Energy level scheme of a three-level atom in free space, interacting with a strong control laser with Rabi frequency $\Omega_\text{con}$ detuned by
    $\Delta_\text{con}$ from the atomic $|g_1\rangle\leftrightarrow|e\rangle$ transition. The inset shows the absorption of a weak probe beam with Rabi frequency
    $\Omega_\text{p}$ as a function of its detuning $\Delta_\text{p}$ from the atomic $|g_2\rangle\leftrightarrow|e\rangle$
    transition, highlighting the region around the state $|+\rangle$, relevant for our experiment. In our measurement, we use a cavity field as a probe field.}
    \label{scheme}
\end{figure}

The optical control mechanism is based on an interference
phenomenon: In a medium with three internal states (Fig.\
\ref{scheme}), a control field $\Omega_\text{con}$ couples a
short-lived excited state $|e\rangle$ to a long-lived ground state
$|g_1\rangle$. Two new eigenstates $|\pm\rangle$, so called
dressed states, are formed, represented by the two absorption
peaks. At two-photon resonance
$(\delta=\Delta_\text{p}-\Delta_\text{con}=0)$, where
$\Delta_\text{p}$ and $\Delta_\text{con}$ are the detunings of
probe and control laser field from the atomic transitions, the
medium becomes transparent due to a destructive interference
between the excitation pathways to the states $|+\rangle$ and
$|-\rangle$. In the case of large detunings
$\Delta_\text{p},\Delta_\text{con}$ compared to the natural line
width $\gamma$, the absorption peaks become strongly asymmetric
due to their very different contributions from states
$|g_1\rangle$ and $|e\rangle$. Here, we investigate a small region
(highlighted in Fig.\ \ref{scheme}) containing both the narrow
state $|+\rangle$ and the transparency point, where both
absorption and dispersion change rapidly with $\delta$, providing
a powerful tool to control the optical properties of an atom.

\begin{figure}

\includegraphics[width=0.45\textwidth]{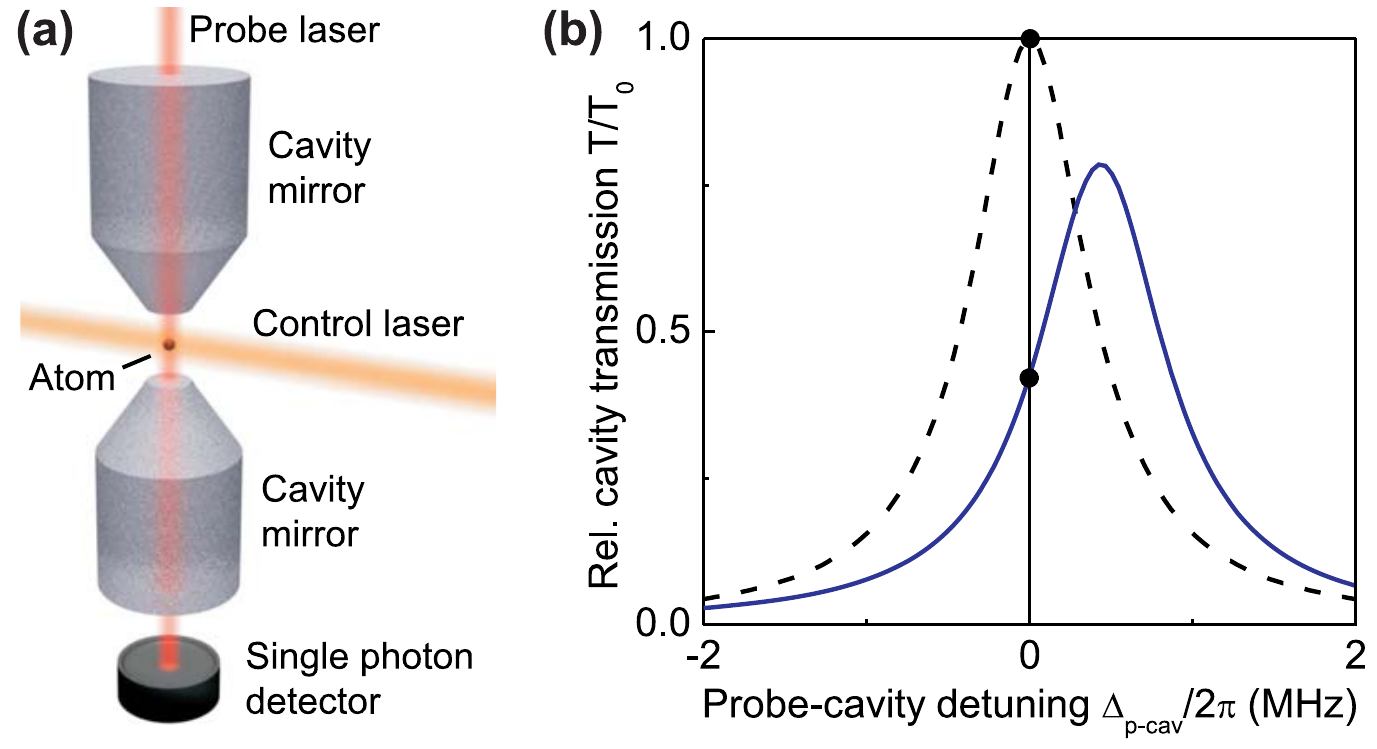}

    \caption{(a) Schematic experimental setup. A single atom is placed between the mirrors of a
     high-finesse cavity and illuminated by a control laser. The transmission of a weak probe laser is
     detected by a single photon counting module.
     (b) Calculated relative cavity transmission $T/T_0$ for zero (dashed line) and one atom in state $|g_2\rangle$ (coupling strength: $g/2\pi=3\text{ MHz}$; no EIT) inside the cavity
     (solid line) as a function of the detuning $\Delta_\text{p-cav}$ of the probe laser from the resonance frequency of the empty cavity.
       The dispersion and absorption by the atom lead to a shift of the cavity resonance and a reduction of
       the maximum transmission, respectively.}
    \label{setup}
\end{figure}

\begin{figure}

\includegraphics[width=0.45\textwidth]{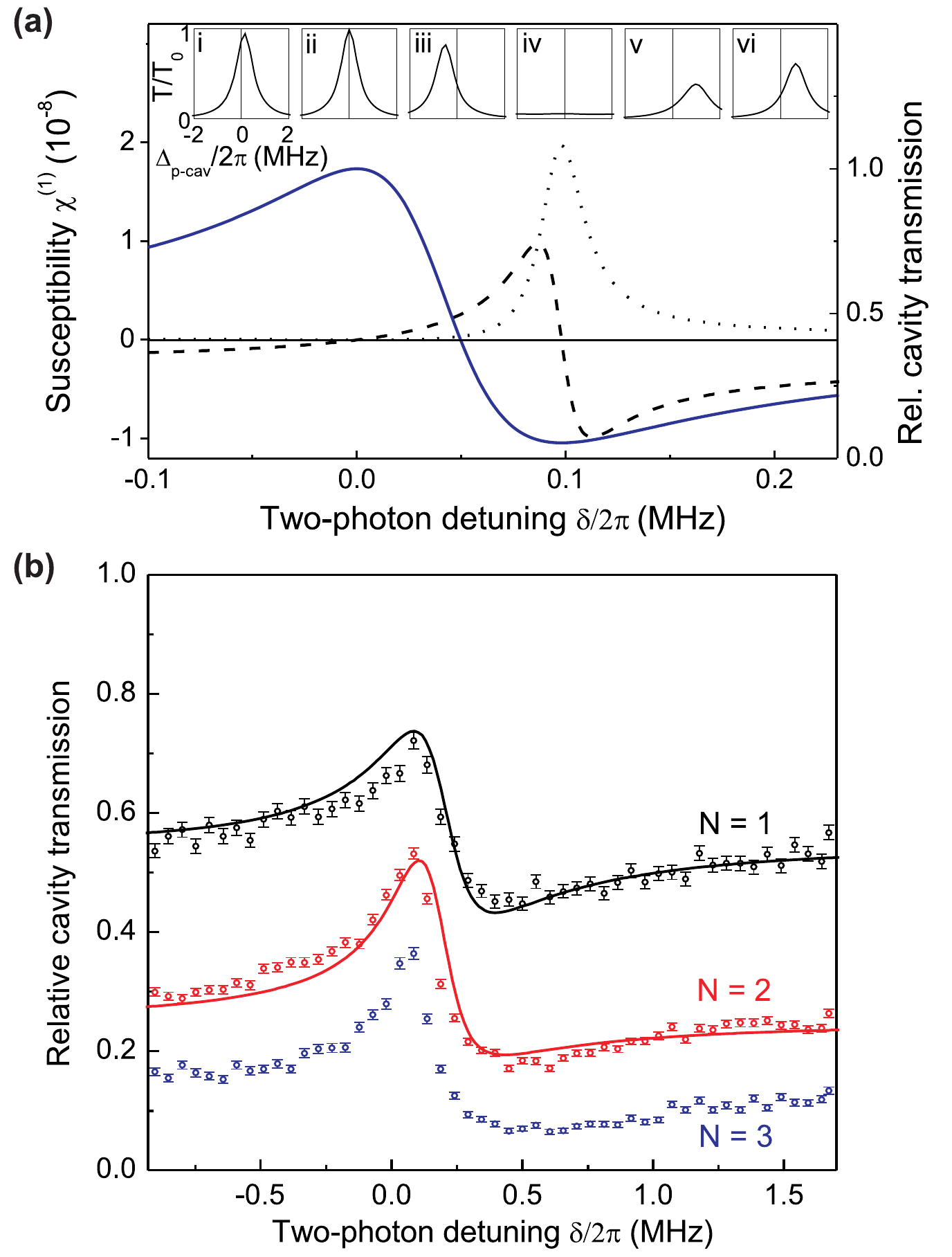}

    \caption{(a) Real part (dashed line) and imaginary part (dotted line) of the
        linear susceptibility $\chi^{(1)}$ of a three-level atom inside the mode volume of a cavity as a function of the two-photon
        detuning $\delta$ with $(\Omega_\text{con},\Delta_\text{p})/2\pi=(2.8,20)\text{ MHz}$. The dotted line corresponds to the absorption inside the highlighted region in Fig.\ \ref{scheme}. The solid line shows the relative probe laser
        transmission for a single atom in the weak probing limit without dephasing, obtained from a numerical solution of the master equation with $g/2\pi=3\text{ MHz}$.
        The small insets (\textsf{i--vi}) are calculated cavity transmission functions as in Fig.\ \ref{setup}(b) for a series of fixed
        values of $\delta$, i.e.\ fixed values of $\chi^{(1)}(\delta)$, with $-50\textrm{ kHz}\leq\delta/2\pi\leq 200\textrm{ kHz}$. (b) Relative transmission of the probe laser through the cavity as a
        function of the two-photon detuning $\delta$, for $N=1,2,3$ atoms. The experimental data are shown as open circles with statistical error.
        The solid lines show numerical solutions of the master equation, taking AC-Stark shifts, dephasing, and the nonvanishing probe laser power into account.
        }
    \label{exp}
\end{figure}

In our experiment we use a single or up to three laser-cooled
cesium atoms where the re\-levant states are the hyperfine ground
states $|g_{1,2}\rangle=|6^2\text{S}_{1/2},F=3,4\rangle$ and the
excited state $|e\rangle=|6^2\text{P}_{3/2},F=4\rangle$ of the
$\text{D}_2$-transition at $852\text{ nm}$. We position our atoms
into a high-finesse cavity ($\mathcal{F}=1.2\cdot10^6$) using an
optical conveyor belt at $1030\text{ nm}$, which also serves as a
permanent trap for the atoms (for details see \cite{Khu08}). Atoms
in state $|g_2\rangle$ are coupled to the resonator field with a
coupling strength $g/2\pi=0\ldots 12\text{ MHz}$, depending on
their Zeeman substate $|m_F|=0,\ldots,4$. The decay rates of the
cavity field and the atomic dipole are $\kappa/2\pi=0.4\text{
MHz}$ and $\gamma/2\pi=2.6\text{ MHz}$, respectively. For strong
coupling, a photon has a high probability to be either absorbed or
experience a significant phase shift already by a single atom. To
change the optical properties of the atoms, we irradiate them with
a control laser beam with a typical power of $1\,\mu\text{W}$ and
a beam diameter of $170\,\mu\text{m}$ propagating orthogonal to
the cavity field, see Fig.\ \ref{setup}(a). A probe laser beam,
resonant with the cavity, populates the empty cavity with on
average $n_\text{p}=0.1$ photons, and the transmitted ones are
detected by a single photon counting module. In order to be
sensitive to both absorptive and dispersive effects and to obtain
a stable atom-cavity coupling, we blue-detune both the probe laser
and the cavity by $\Delta_\text{p}/2\pi = 20\text{ MHz}$ from the
atomic $|g_2\rangle\leftrightarrow|e\rangle$ transition. Thereby,
we simultaneously use the cavity as a signal amplifier and as a
phase-to-amplitude converter, see Fig.\ \ref{setup}(b): The
dispersive effect of a single atom in state $|g_2\rangle$ shifts
the cavity resonance, while the absorption directly reduces the
overall amplitude, such that the transmission of the probe, now
situated on the steep slope, drops to about $50\%$ of the empty
cavity level.

We record the cavity transmission while sweeping the two-photon
detuning within $10\text{ ms}$ from $\delta= -0.9\text{ MHz}$ to
$1.7 \text{ MHz}$ back and forth after $N=1,2,3$ atoms have been
placed into the cavity, see Fig.\ \ref{exp}(b). We always keep the
probe-cavity detuning $\Delta_\text{p-cav}=0$, different from
normal-mode scans \cite{Muc10}, and scan $\Delta_\text{con}$
instead. We alternate each sweep with a $10\text{ ms}$ cooling
interval by setting $\delta=0$ (see below). After 20 measurement
and cooling intervals, we switch the control beam off and a
repumper on, resonant with the $|g_1\rangle \leftrightarrow
|e\rangle$ transition, and record the cavity transmission for
another $20\text{ ms}$. Then, the atoms are retrieved from the
cavity and counted again. We post-select those data traces in
which no atom has been lost and all atoms have coupled strongly to
the cavity mode: We first check if the relative cavity
transmission at the end of each sequence is below 70\%. For more
than one atom, we also analyze a fluorescence image \cite{Mir03}
taken at the beginning of each sequence and check if the atoms are
well within the cavity mode, i.e.\ that their average
position-dependent coupling is at least 95\% of the maximum value.
Each data trace shown in Fig.\ \ref{exp}(b) is an average of about
100 single sequences. We observe a strong dispersive EIT signal
close to the two-photon resonance: At $\delta\approx 0$, we find a
transmission maximum followed by a minimum about $250\text{ kHz}$
away.

This shape of the cavity transmission signal, which is determined
by both absorption and dispersion of the atoms, can be
qualitatively understood with the help of a semiclassical model
\cite{Fle05}: Fig.\ \ref{exp}(a) shows the real (dispersive) and
imaginary (absorptive) part of the linear susceptibility
$\chi^{(1)}$ of an atom with three internal states, off-resonantly
interacting with a strong control and a weak probe field. For the
case of one atom inside an optical cavity, the calculated
transmission in the limit of weak probing is shown as a solid
line. Here we consider the atom coupled with $g/2\pi=3.0\text{
MHz}$ to the cavity mode and the Rabi frequency of the control
field of $\Omega_\text{con}/2\pi=2.8\text{ MHz}$, as obtained from
the model (see below).

\begin{figure}

\includegraphics[width=0.45\textwidth]{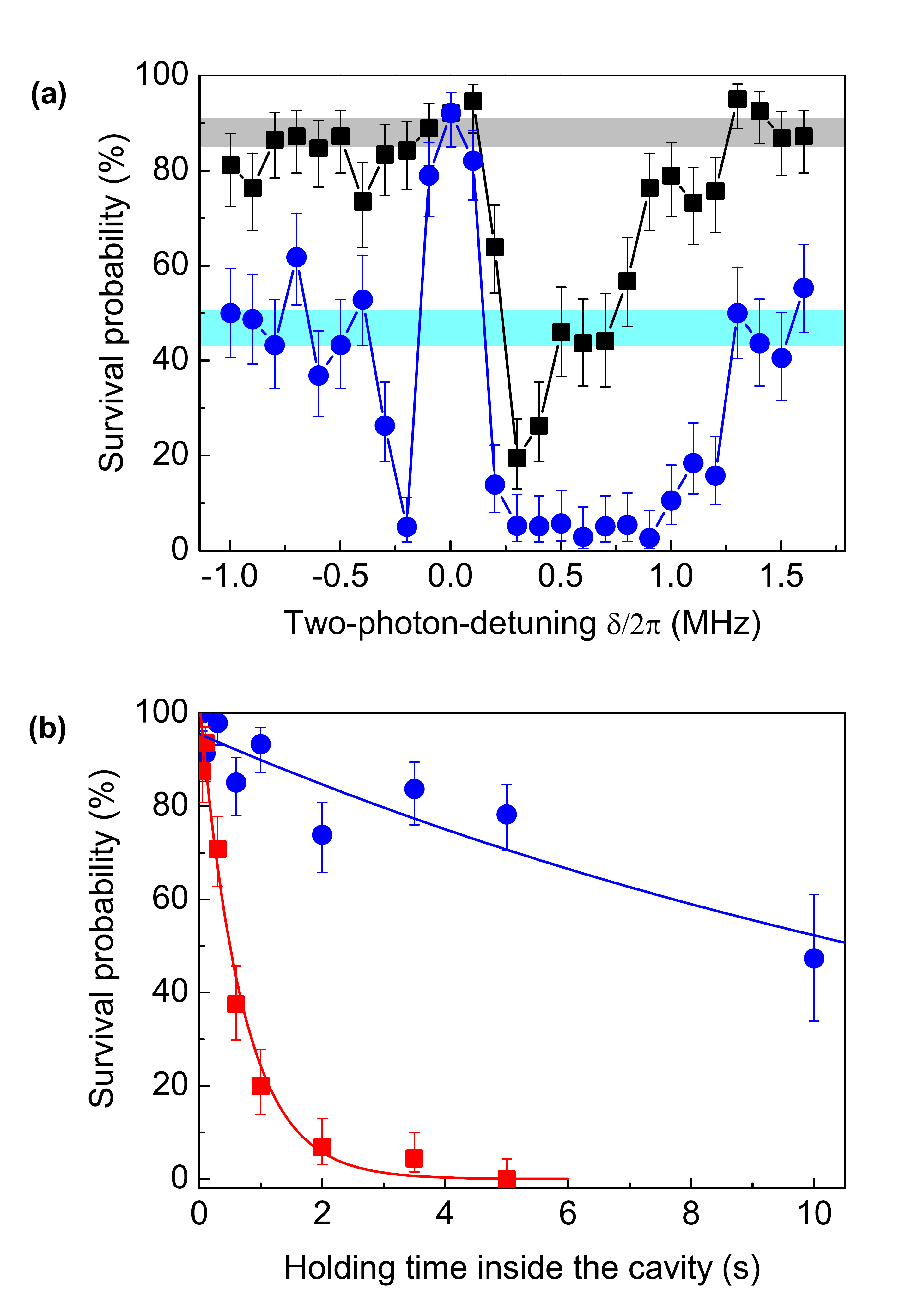}

    \caption{(a) Probability of one atom to remain trapped inside the cavity (survival probability) as a function of the two-photon
     detuning $\delta$ for a holding time of $20\text{ ms}$ (squares) and $300\text{ ms}$ (circles).
     The shaded regions indicate the corresponding probabilities (top to bottom) for large two-photon detunings.
       (b) Survival probability of single atoms inside the cavity as a function of time.
     The circles show the survival probability with the control laser at two-photon resonance and the squares show the survival with
      the far-detuned trapping lasers switched on only. The error bars in both graphs are statistical.}
    \label{cooling}
\end{figure}

Depending on the two-photon detuning $\delta$, we identify three different
regimes:

At two-photon resonance $\chi^{(1)}$ vanishes in the ideal case
without dephasing [Fig.\ \ref{exp}(a), region \textsf{ii}], thus
the medium becomes transparent and the refractive index takes the
vacuum value $n=(1+\chi^{(1)})^{1/2}=1$. This leads to a cavity
transmission equal to the empty cavity case.

At
$\delta_\text{abs}\approx\Omega_\text{con}^2/(4\Delta_\text{con})\approx
2\pi\times 0.1\text{ MHz}$ (\textsf{iv}), the model shows an
absorption peak. Here, the increased losses from the resonator due
to spontaneous scattering of the probe laser field by the atom
lead to a transmission minimum. The full width at half maximum of
the absorption peak
$\Delta\delta_\text{abs}\approx\gamma\Omega_\text{con}^2/(2\Delta_\text{con}^2)\approx
2\pi\times 25\text{ kHz}$ is much less than the atomic line width
$\gamma$, illustrating the large ground state contribution to the
dressed state $|+\rangle = (1-\epsilon^2)^{1/2}|g_1\rangle+
\epsilon |e\rangle$, with
$\epsilon\approx\Omega_\text{con}/(2\Delta_\text{con})\ll1$
\footnote{The 25 kHz linewidth is also the scattering rate of the
control light for atoms in state $\left|g_1\right>$.}.

In the other regions (\textsf{i, iii, v, vi}), dispersion
dominates over absorption. Here, the change in transmission is
caused mainly by the shift of the cavity resonance frequency,
tuning the cavity out of resonance with the probe laser. The sign
of dispersion changes twice, at $\delta=0$ and
$\delta=\delta_\text{abs}$. In the regions (\textsf{i, v, vi})
with negative dispersion and thus $\text{Re}(n)<1$, the cavity
resonance is shifted to larger frequencies, and vice versa.
Finally, in the limit of large two-photon detunings
$|\delta|\gg\delta_\text{abs}$, the control laser essentially acts
as an incoherent repumping laser, which pumps atoms that are in
state $|g_1\rangle$ to state $|g_2\rangle$, and the transmission
approaches the value for a single atom in $|g_2\rangle$, as in
Fig.\ \ref{setup}(b).

For our experimental system, one has to take into account the
deviation from the weak-probing limit due to the nonzero probe
laser power (in a classical picture: $\Omega_\text{p}=
2g\sqrt{n_\text{p}}\leq2\pi\times 2\text{ MHz}\approx
\Omega_\text{con}$), as well as the contributions of other excited
hyperfine states. In addition, ground state dephasing, caused by
spatial and temporal inhomogeneities, e.g.\, fluctuating
atom-cavity coupling, inhomogeneous light shifts and residual
magnetic fields has been included in a more realistic model. The
phenomenological ground state dephasing rate $\gamma_\text{deph}$
sets a lower limit for the width of the slope of the EIT signal
and reduces the maximum transmission and increases the minimum
transmission. We calculate the cavity transmission as a function
of the two-photon detuning $\delta$ by finding the steady-state
density matrix. We approximate the system by $N$ atomic five-level
systems with two ground states
$|g_{1,2}\rangle=|6^2\text{S}_{1/2},F=3,4\rangle$ and three
excited states $|d\rangle,|e\rangle,|f\rangle =
|6^2\text{P}_{3/2},F=3,4,5\rangle$. The ground state $|g_2\rangle$
is coupled to all three excited states via the cavity field, and
ground state $|g_1\rangle$ is coupled to $|d\rangle$ and
$|e\rangle$ via the control laser ($|g_1\rangle \leftrightarrow
|f\rangle$ is forbidden by selection rules). The dissipative
processes, like photon decay from the cavity or population decay
from the excited states as well as ground state dephasing, are
included into the description using a master equation. For
computational reasons, we further restrict the number of Fock
states of the cavity field to three (corresponding to photon
numbers $n_\text{photon}=0,1,2$). As the dimension of the Hilbert
space scales exponentially with the number of atoms $N$, only the
one and two atom cases have been calculated numerically. This
model [Fig.\ \ref{exp}(b)] fits the data with the same set of
parameters
$(g,\Omega_\text{con},\gamma_\text{deph})/2\pi=(3.0,2.8,0.15)\text{
MHz}$ for one and two atoms, except for an independently measured,
additional differential light shift of approx. $100\text{ kHz}$
originating mainly from the control laser. The effective value of
$g$ arises mainly from the distribution over Zeeman sub-states but
also from motional averaging.

So far, only the dynamics of the internal states has been
discussed. However, the interaction of atoms with near-resonant
light fields is also strongly connected with their motional state
due to the exchange of photon momenta. The interaction of
two-level atoms with the cavity field can induce strong cooling
forces \cite{Dom03}, leading to, e.g.\, long storage times in
experiments \cite{Nus05}. Cavity cooling is probably of relevance
also for our system \cite{Khu08}. For three-level atoms, the
creation of a transparency window close to an absorption peak much
narrower than the atomic line width by EIT can give rise to
another sub-Doppler cooling mechanism \cite{Mor00}, as has been
demonstrated with trapped ions \cite{Roo00,Sch01}. In our
experiment, strong cooling and heating effects close to the
two-photon resonance have been observed. We have measured the
probability of one atom to remain trapped inside the cavity for a
time interval of $20\text{ ms}$ and $300\text{ ms}$, respectively,
for different values of the two-photon detuning $\delta$ and have
compared it with the survival probabilities for large two-photon
detunings, see Fig.\ \ref{cooling}(a). We have observed a cooling
region at $\delta=0$ in between a narrow weak and a strong broad
asymmetric heating region at $\delta/2\pi=-0.2\text{ MHz}$ and
$\delta/2\pi=0.3\text{ MHz}$, respectively. Without cooling, i.e.\
only with the far-off resonant trapping lasers, the $1/e$ lifetime
of atoms inside the resonator is $(0.7\pm0.1)\text{ s}$, see Fig.\
\ref{cooling}(b). The cooling at $\delta=0$ extends the lifetime
to $(16\pm3)\text{ s}$. Far from the two-photon-resonance, the
lifetime is compatible with the value of the uncooled atoms. While
a clean distinction between cavity and EIT-cooling effects would
require an advanced theoretical model, we utilize this cooling
mechanism in our experimental sequence for the measurement of the
EIT spectrum. By alternating cooling intervals with measurement
intervals, we are able to increase the number of measurement
cycles per atom by about a factor of 5. The cooling mechanism
could be further employed to obtain a stronger and more stable
coupling of single particles to a cavity as is desired for many
protocols in quantum information processing.

Our demonstration of optical control of the optical properties of a single atom
has numerous implications for quantum engineering with single atoms: While
linear absorption is suppressed due to destructive interference in EIT,
nonlinear susceptibilities, giving rise to frequency summing and parametric
amplification, can be resonantly enhanced \cite{Har90,Yan01}. Furthermore,
nonlinearities lead to an effective interaction between single photons. This is
the basis of quantum logic gates \cite{Tur95,Dua04}. However, in our case,
still many photons are needed in the control laser beam to change the optical
properties of the atom. Using a second cavity mode to enhance the interaction
with the control beam could lead to an optical transistor for single photons
\cite{Sol05,Ber06}.

\begin{acknowledgments}
We would like to thank Lingbo Kong and Jonathan Simon for early
ideas. We acknowledge financial support by the EC through AQUTE.
R.R. and T.K acknowledge support from the Studienstiftung des
deutschen Volkes and the Bonn-Cologne Graduate School of Physics
and Astronomy.
\end{acknowledgments}

%

\end{document}